\documentclass[twocolumn,showpacs,superscriptaddress,preprintnumbers,amsmath,amssymb]{revtex4}

\usepackage{epsf}
\usepackage{graphicx}

\begin{document}

%%\draft  % \draft command makes pacs numbers print

\title{A lattice study of the masses of singlet $0^{++}$ mesons}

\author{A. Hart}
\affiliation{
SUPA, School of Physics, University of Edinburgh,
Edinburgh EH9~3JZ, Scotland
}
\author{C. McNeile}
\thanks{Now at the Department of Physics and Astronomy, University of Glasgow.}
\affiliation{
Theoretical Physics Division, Dept. of Mathematical Sciences, 
          University of Liverpool, Liverpool L69 3BX, UK}
\author{C.~Michael}
\affiliation{
Theoretical Physics Division, Dept. of Mathematical Sciences, 
          University of Liverpool, Liverpool L69 3BX, UK}

\author{J.~Pickavance}
\affiliation{
Theoretical Physics Division, Dept. of Mathematical Sciences, 
          University of Liverpool, Liverpool L69 3BX, UK}

\collaboration{UKQCD Collaboration}
\noaffiliation

\date{\today}

\begin{abstract}
We compute the masses of the flavour singlet $0^{++}$ mesons using
($n_f=2$) unquenched lattice QCD with the Iwasaki and Wilson gauge actions.
Both fermionic and glueball interpolating operators are used
to create the states. The mass of the lightest $0^{++}$ meson
is suppressed relative to the mass of the $0^{++}$ glueball
in quenched QCD at an equivalent lattice spacing.
We discuss two possible physical reasons for this.
\end{abstract}

% Lattice QCD Calculations
%%\pacs{12.38.Gc} 

\maketitle

% body of paper here

\section{Introduction}

The interpretation of the experimental  $f_0$ mesons in terms of
fundamental quark and  glue fields is still not 
settled~\cite{Klempt:2000ud,Close:2001zp,Barnes:2002pw,Close:2002zu,Close:2004ip,Pennington:2005be}. 
Quenched
lattice QCD predicts that the mass of the scalar ($J^{PC}=0^{++}$)
glueball  is around
1.6~GeV~\cite{Bali:1993fb,Morningstar:1997ff,Morningstar:1999rf,Chen:2005mg}.
 Hence, attention has focused on finding evidence for a 
glueball component (mixing with quark-antiquark) in the physical
$f_0(1370)$, $f_0(1500)$, and $f_0(1710)$ mesons. The quark model
predicts that there are two $f_0$ mesons in this mass regime, so the 
existence of  three mesons is suggestive of the presence of  additional
degrees of freedom such as the elusive $0^{++}$ glueball.

The experimental $f_0$ spectrum contains more puzzles. The
experimental data for the $f_0(1370)$ still seems 
controversial~\cite{Close:2001zp}. 
There
is also a new state $f_0(1790)$ reported by
BES~\cite{Ablikim:2006dw,Close:2004ip}.  It is not clear how this
effects the ``standard'' mixing scenario for the $0^{++}$ mesons, that
consider only three mesons.  The interpretation of the $f_0(980)$
meson which is close to the KK threshold,
 in terms of quark and antiquarks, is also uncertain. Even extracting
the mass and width of the $f_0$(400-1200) state is still
controversial~\cite{Close:2001zp,Barnes:2002pw}, but progress seems to
have been made recently~\cite{Caprini:2005zr,Leutwyler:2006gz}.
It
may be difficult for lattice calculations to explore the
$f_0$(400-1200), because it has such a large width. Although there are
attempts to study this state~\cite{Kunihiro:2003yj,Mathur:2006bs} on
the lattice.

%%  big decay width
%%  does quenched QCD have anything to do with reality
%%

\begin{figure}[t]
\begin{center}
\leavevmode
\includegraphics[scale=0.4,clip]{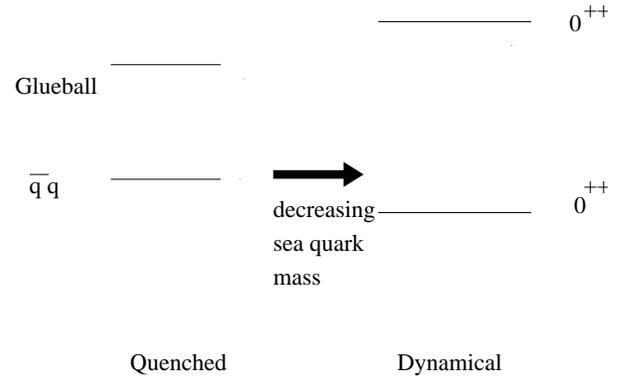}
\end{center}
\caption[]{\label{fig:Mixing} {
Mixing as unquenching occurs. 
}}
\end{figure}

The glueball spectrum in pure gauge theory is theoretically well
defined, because the glueball operators do not mix with  fermionic
$\overline{\psi}\psi$ operators. The work of many authors has shown that
the lightest $0^{++}$ state is at 1640(40)MeV (plus errors of around
10\% in setting the scale)~\cite{Teper:1998kw,Bali:1993fb}  in pure SU(3)
Yang-Mills theory.  In unquenched calculations there are new
complications. The glueball and fermionic $0^{++}$ states can mix
as illustrated in figure~\ref{fig:Mixing}. In fact, it no longer
makes sense to talk about glueball states in dynamical QCD calculations
(although there are glueball interpolating operators). There
are only flavour singlet $0^{++}$ mesons that we sometimes denote
by FS.
Also
the $0^{++}$ states can decay into meson pairs, so the decay width of
hadron may play an important  role in the dynamics.

 Unlike in quenched QCD which has a degenerate isoscalar and  isovector
scalar meson (made from $\bar{q} q$), the isoscalar state mixes  with the
glueball, being reduced in mass if it is lighter than the glueball.
 In detail, this mixing will be a function of 
the lattice parameters and sea quark masses. These will have to be 
extrapolated to their physical values to get the physical
mixing. 

There are predictions~\cite{Narison:2000dh,Pennington:1998ys} 
that the width for glueball
decay to two mesons is large relative to its mass.  Although one
exploratory lattice calculation~\cite{Sexton:1995kd} found that the
decay width of the  $0^{++}$ was 108(29) MeV in the quenched
approximation.

Indeed, the standard formalism for determining the masses of hadrons on
a lattice may be inappropriate in the presence of open decay channels 
and special  techniques may be
needed~\cite{Michael:1989mf,DeGrand:1991ip,McNeile:2003dy}. The MILC
collaboration reported problems in extracting the masses of non-singlet
$0^{++}$~\cite{Bernard:2001av}  and $1^{-+}$~\cite{Bernard:2003jd}
mesons, that they attributed to open decay channels. The techniques to
study decay widths using lattice QCD have recently been
reviewed~\cite{Michael:2005kw}. All the above discussions   indicate
that it is essential to  study the singlet $0^{++}$ mass  spectrum with
dynamical fermions. 

%% New analysis: finer lattices
%%               nonzero momentum
%%   nonperturbatively improved
%%

 The basic formalism for the mixing of pure glue and $\overline{q}q$
states was described in previous work~\cite{McNeile:2000xx} from
UKQCD.  That paper obtained surprisingly light values  for the
flavour-singlet scalar meson, but at relatively coarse lattice spacing.
 Here we explore this issue using a smaller lattice spacing. We also
determine  the spectrum using lattices with a different gauge action
since the  lattice artifacts (for instance big order $a^2$ corrections
to the scalar mass) will then be different.

In detail, we use dynamical gauge configurations generated by UKQCD
using fully non-perturbatively improved clover
fermions~\cite{Allton:2001sk}. The lattice spacing is finer, $a \sim
0.1$~fm, than in our previous study~\cite{McNeile:2000xx}.  Another
technical improvement is that we use fermionic scalar mesons at nonzero
momentum.  This helps to get a better signal for the larger volumes.
This was regularly used  for pure glue operators~\cite{Hart:2001fp}, but
is not routinely used  for fermionic operators. The glueball spectrum
using only glue operators has been presented
in~\cite{Allton:2001sk,Hart:2001fp}. The inclusion of $\overline{q}q$
operators in this study allows us to address mixing issues. UKQCD has
recently published a lattice study of the non-singlet $0^{++}$
mesons~\cite{McNeile:2006nv}.

We also use gauge configurations from the CP-PACS
collaboration~\cite{AliKhan:2001tx}. These configurations were
generated with an improved gauge action. This can help to address some
of the potential issues with using the clover action in combination
with the Wilson pure gauge action~\cite{Hart:2001tv,Sommer:2003ne}. In particular
the lattice spacing dependence of the mass of the $0^{++}$ states is
known to be large with the Wilson pure gauge
action~\cite{Morningstar:1997ff,Morningstar:1999rf,Morningstar:1998du,Necco:2003vh}.
There have also been claims that the unquenched calculations that use
the clover fermion action, with the Wilson single plaquette action,
 are 
affected by bulk phase transition in the 
adjoint plane~\cite{Hart:2001tv,Aoki:2004iq,Jansen:2003nt,Sommer:2003ne,Necco:2003vh,Farchioni:2004fs}. 
One conjectured consequence of the phase transition was the 
suppression of the $0^{++}$ glueball masses obtained
by Hart and Teper~\cite{Hart:2001fp} from $N_f=2$ unquenched
QCD. The JLQCD collaboration~\cite{Aoki:2004iq}  found the effect
of the adjoint phase transition was reduced by the use of improved
gauge actions, such as the Iwasaki action. Hence the CP-PACS 
data will be an important cross check on results.

\section{Parameters of the lattice calculation}

We use gauge configurations from the 
UKQCD~\cite{Allton:2001sk} 
and CP-PACS~\cite{AliKhan:2001tx}
collaborations.
For the gauge configurations from UKQCD, 
the non-perturbatively improved
clover action was used to generate the 
unquenched gauge configurations,
with a clover coefficient $c_{SW} = 2.0171$.  
The $\beta$ value was 5.2 
and the lattice
volume was $16^3$ x $32$.
The sea quark $\kappa$ values of 0.1350 and 0.1355
were used for the singlet 
correlators.
The hadron spectrum, potential, and some glueball 
estimates from this data set have been
presented in~\cite{Allton:2001sk,Hart:2001fp}.

We use two sets of data with $N_f=2$ from the CP-PACS collaboration at 
$\beta=1.95$~\cite{AliKhan:2001tx}. 
The tadpole improved clover action with clover
coefficient of $c_{SW}=1.53$.  The Iwasaki renormalised group improved
gauge action was used.
The lattice size is $16^3$ x $32$ and we use valence quark 
masses
equal to those of the sea quarks.  The lattice details are summarised
in Table~\ref{sim}. 
\vspace{0.5cm}
\begin{table}[htbp]
\begin{center}
\begin{tabular}{|c|c|c|c|c|}
\hline
Code & $N_{gauge}$ & $\kappa$ & $r_0/a$ & $am_{PS}$ \\
\hline
C390 & 648 & 0.1390 & 2.651 & 0.729 \\
C410 & 490 & 0.1410 & 3.014 & 0.427 \\
U350 & 144 & 0.1350 & 4.75  & 0.405  \\
U355 & 416 & 0.1355 & 5.04  & 0.294  \\
\hline
\end{tabular}
\caption{Simulation details for CP-PACS and UKQCD data sets.}
\label{sim}
\end{center}
\end{table}

The methods we use to compute the disconnected diagrams
have been described in earlier 
publications~\cite{McNeile:2000xx,Allton:2004qq}.
Essentially we use random $Z_2$ volume sources
to estimate the bubble diagrams, with the variance
reduction technique described in~\cite{McNeile:2000xx}.
We have also published data for the 
singlet pseudoscalar channel~\cite{Allton:2004qq}
from some of this data set. 
We have presented preliminary results for a subset of the 
data used in this paper in~\cite{Hart:2002sp}.

\section{Fit methods}

To extract a good signal from lattice QCD calculations,
it has been found to be essential to use 
a variational basis of correlators.
\begin{equation}
C_{i \; j}(t)  = \langle 0 \mid {\cal O}_i(t)^{\dagger}
                         {\cal O}_j(0) \mid 0   \rangle
\label{eq:vartMAT}
\end{equation}
In terms of the path integral the matrix $C_{ij}$
is
%%%
\begin{equation}
C_{i \; j}(t)  = \frac{1}{{\cal Z}}
\int dU
\int d\psi  
 \int d\overline{\psi}
e^{-S_F - S_G}
{\cal O}_i(t)^{\dagger}
{\cal O}_j(0) 
\end{equation}
where $S_F$ is the clover fermion action
(lattice approximation to the Dirac Lagrangian)
and $S_G$ is Wilson gauge action (lattice approximation to the
gauge action). The full details of the Lagrangians are described in
our earlier work on the light hadron spectrum~\cite{Allton:2001sk}.
In the fermion sector we use fuzzed and local operators as basis
states ($O_i(t)$). The fuzzing method is described in the spectrum
study~\cite{Allton:2001sk} of UKQCD. In the pure glue sector, two types of
smeared glueball operators~\cite{Hart:2001fp} are included.

We use factorising (or variational) fits to extract the  masses and
amplitudes. 
 \begin{equation} C_{i \; j}(t)  = \sum_{m=0}^{M} 
c_{i}^{m} c_{j}^{m} e^{- E_m t } \label{eq:matrixFIT}
 \end{equation} 
 The scalar channel $(0^{++})$ at zero momentum has a vacuum
contribution which we account for  either by fixing  $E_0 =
0$~\cite{Niedermayer:2000yx} or by evaluating vacuum subtracted
correlations. $M$  is the  number of states in the fit  which can be 1
to 3.   For our multi-state fits, we quote the masses for all states.
However we regard the highest state as a ``truncation error'', and this
mass may not correspond to a  physical state.

Any state with the same quantum numbers as the  operator ${\cal O}_j$
will couple to that channel. If the amplitude is small then it will be
difficult to extract the mass of the state. We feel it is better to use
as many different interpolating operators in constructing the correlator
matrix. If the singlet $0^{++}$ states couple to both pure glue and
fermionic operators, then it must be better to  use  additional basis
states, provided  they are not too noisy and are sufficiently
independent of the other states. This  can help to stablize the
multi-state fits. Hence our best results are from fits from order 4
smearing matrices that include both glueball and  fermionic operators,
each with two spatial sizes.

The Iwasaki action has ghost states that contribute to the correlators
at small time  distances~\cite{Luscher:1984is,Necco:2003vh}. This issue
has recently been studied by Necco~\cite{Necco:2003vh}  in quenched QCD
for a variety of improved gauge actions. Here we shall restrict our 
fits to $t > a$ which should reduce these potential problems.

% The identification of physical mesons with  states on the  lattice
%requires a careful extrapolation in quark mass.

\section{Results}

\begin{figure}
%\vbox{\epsfxsize=4in \epsfbox{meff_scalar_p100+010+001.ps}}
\vbox{\epsfxsize=4in \epsfbox{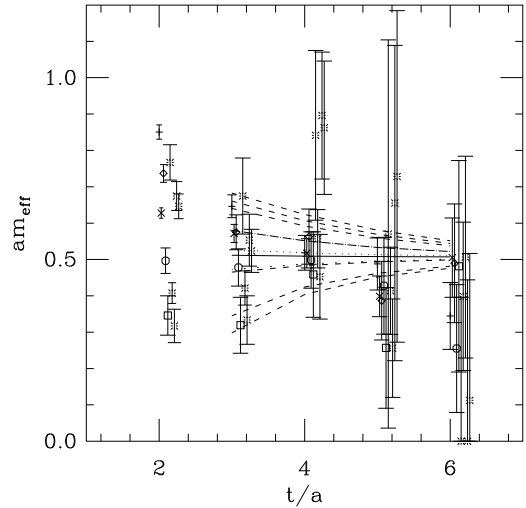}}
%%\vskip -0.5in
%%\vskip -9mm
 \caption{Singlet scalar meson effective mass plot from fermionic and
gluonic operators from U355 with momentum zero.
 }
\label{fig:Smeffscalar}
%%\vskip -8mm
\end{figure}

In figure~\ref{fig:Smeffscalar} we show an effective
mass plot for the U355 data at zero momentum together with a fit.
This illustrates the quality of the data and also the advantage 
of having many different correlations to fit simultaneously.

\begin{table}[tb]
  \caption{Fit results for flavour singlet scalar mesons.  These are
fits to four by four matrices of correlations to the $t$-range shown
with momentum ($p=2 \pi n/L$. Fits to momentum zero are to the vacuum
subtracted correlators here. For these fits involving disconnected
contributions,  we expect a $\chi^2$/dof  close to 1.0 since the
absolute error stays constant with increasing $t$. 
 }
  \begin{tabular}{cccc|cccc}
Code  & n  & Region  & $\chi^2$/dof & $E_0$ & $E_1$& $E_2$
\\ \hline
 %tmsc202 fits
U350   &  000  & 2 -- 6  &  24/40  &  0.64(3)& 1.17(25)& -\\
U350   &  100  & 2 -- 6  & 24/35   & 0.69(4)& 0.86(7)& 1.23(29) \\
 %tm0016 fits mom=0
U355   &  000  & 2 -- 6  &  24/40  & 0.510(14)&1.06(12) & - \\
 % tmsc202 (see tmf0202 for mom=/= 0 355 also: more stats?)
%U355   &  100  & 2 -- 6  & 20/35   &  0.64(3)& 0.94(6)& 1.41(27)\\
   %  tmf0202               
U355   &  100  & 2 -- 6  & 34/35   &  0.62(3)& 1.01(5)& 1.47(10)\\
 % convert to mass                         .48      .93  
% U355   &  110  & 2 -- 6  & 40/35   & 0.78(3)& 0.96(4)& 1.51(18)\\
   % tmf0202                    36/35      .75 2   1.05 3  1.63 9
 U355   &  110  & 2 -- 6  & 36/35   & 0.75(2)& 1.05(3)& 1.63(9)\\
 % convert to mass                         .50      .89  
 % 55 averages over mom                      .51(2)   .97(5)
C390   &  000  & 2 -- 6  & 45/45   & 0.98(4) &-&-\\
%C390 & FS  &  000  & 2 -- 6  &   0.27   & 0.944(+70/-110)  \\
%C410 & FS  &  000  & 2 -- 6  &   0.46   & 0.669(20)  \\
C410  &  000  & 2 -- 6  & 19/40   & 0.67(2) & 1.56(40) &- \\
%C390 & GB  &  000  & 2 -- 6  &  0.19    & 0.994(+11/-10)  \\
%C410 & GB  &  000  & 2 -- 6  &   0.70   & 0.729(+46/-49) \\
\hline
  \end{tabular}
 \label{tab:FIT4x4RESULTS}
 \end{table}

In Table~\ref{tab:FIT4x4RESULTS}, we collect our fit results for the 
flavour singlet scalar. Results for the flavour non-singlet 
were presented previously~\cite{McNeile:2006nv}. 
These  results use the 4 by 4 matrix of correlators that includes
pure glue and pure fermionic operators in the same fit. 
 For connected correlators we use the conventional 
method with sources at the origin (with 4 time sources for U355).
 For disconnected fermionic loops we use the method of~\cite{McNeile:2000xx} 
with 100 stochastic samples. 
 For \lq\lq glueball\rq\rq operators we used square fuzzed Wilson loops of
differing sizes,  selected to have good overlap with the scalar
glueball~\cite{Hart:2001fp}. 
 We measure  glueball operators every 10 gauge configurations  and
fermionic disconnected operators every 10 (40 for U350 and 20 for 
U355).
  The   data are binned into groups of 40 trajectories to avoid the
impact of any autocorrelation on the error analysis.

We use correlated fits~\cite{Michael:1995sz}  to ascertain the minimum
$t$-value that gives an acceptable fit. We then quote results from an
uncorrelated fit to that range. We find similar results for the two
lightest singlet states at non-zero momentum from 3 state fits to $t$
ranges of 2-6 to  2-10  and from 2 state fits for $t$ ranges of 3-6 to
3-10. Thus the systematic error  on the ground state energy from varying
the fits seems to be smaller than the statistical error which is quoted
in the table.

\begin{figure}
\vbox{\epsfxsize=4in \epsfbox{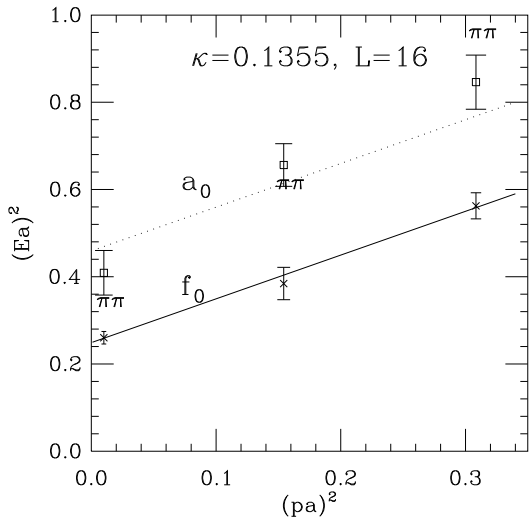}}
 \caption{Scalar meson masses (for U355)  versus momentum for 
both flavour singlet ($f_0$) and non-singlet ($a_0$). The straight 
lines show the expected slope for a relativistic mass-energy relationship.
The lowest $\pi \pi$ thresholds are also illustrated.
 }
\label{fig:mom}

\end{figure}

For the FS channel, the non-zero momentum  results are more stable  
since they  do not involve a vacuum subtraction and also they involve
additional  statistics from constructing the correlator from momenta in each 
spatial direction.
 We expect that
 \begin{equation}
 E^2=m^2+p^2
 \label{eq:adjust}
 \end{equation}
 where $p=2n\pi/L$ with integer vector $n$ where we explore $n.n=0,\ 1,\
2$  here. As shown in  fig.~\ref{fig:mom}, we find consistency (U355 is
plotted since for  that case we have 3 momentum values) within  our
statistical errors.  
The mass value  is then the  intercept
and is consistent with the results in table~\ref{tab:FIT4x4RESULTS}.

Using glueball operators at zero and nonzero
momentum, UKQCD~\cite{Allton:2001sk,Hart:2001fp} 
previously found $a M^{0++}$ = 0.628(30) 
 and $a M^{0++}$ = 0.626(41) 
 at $\kappa = 0.1355$ and at $\kappa = 0.1350$ respectively.
The masses were estimated using effective masses after
finding the optimal basis using a variational 
technique~\cite{Hart:2001fp}. This previous analysis claimed 
that the $0^{++}$ state was lighter in $N_f=2$ with the 
mass ratio from dynamical to quenched lattices of 0.85(3).

Using equation~\ref{eq:adjust} with the data in
table~\ref{tab:FIT4x4RESULTS},
we obtain $a m_1$($0^{++}$) = 0.51(2) for U355 and 0.60(4) for U355. 
Hence, doing a combined fit to glueball and $\overline{q}q$
$0^{++}$ operators has produced additional suppression of the 
$0^{++}$ mass with the inclusion of dynamical fermions.
 It is only for U355 that we have a reliable estimate of the first
excited  state, obtaining $a m_2$($0^{++}$) = 0.92(6). 
 For U355, we find that the fit coefficients of our two lightest states
have  the structure expected from a maximal mixing of a gluonic state and 
a fermionic state.
 The previous analysis~\cite{Allton:2001sk,Hart:2001fp} using only
gluonic operators  appears to have been  getting close to the weighted
mean of the first and second masses which is consistent with the maximal 
mixing we find here.

\begin{figure}[t]
\begin{center}
\leavevmode
\includegraphics[scale=0.35,angle=-90,clip]{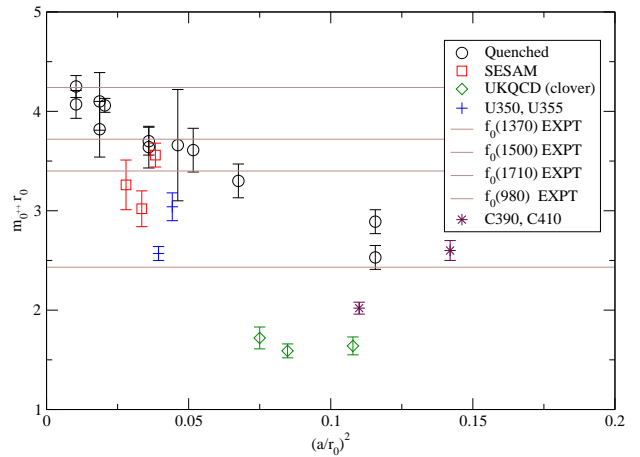}
\end{center}
\caption[]{\label{fig:contlimit} {
Compilation of FS $0^{++}$ masses versus $a^2$.
}}
\end{figure}

 In figure~\ref{fig:contlimit} the $0^{++}$ masses (from the four 
basis fit) are plotted with the continuum limit of the quenched
data (with standard Wilson glue). The masses are in units of 
$r_0$, that is convenient unit for lattice studies.
To convert to MeV the reader can use $1/r_0 \sim $ 400 MeV.

Our study using the CP-PACS lattices with Iwasaki glue has revealed
somewhat similar results to those obtained previously using Wilson
glue~\cite{McNeile:2000xx} at similar  lattice spacings. 
Our results (U355 and U350) with 
finer lattice spacings also illustrate the suppression of the 
mass of the FS $0^{++}$ state relative to the quenched $0^{++}$ glueball
mass at a similar lattice spacing.
Thus the feature found previously~\cite{McNeile:2000xx} of surprisingly
light $f_0$ mass is consistent with the new results at a finer
lattice spacing and with an improved gauge action at the 
coarse action. We discuss this further in section~\ref{se:conclusions}.

We note that in this study the lightest singlet scalar ($f_0$) is not
substantially  lighter than the non-singlet ($a_0$), as shown in 
fig~\ref{fig:mom} for instance. 
In our previous work at $\beta = 5.2$ using the clover action with
$c_{SW}=1.76$~\cite{McNeile:2000xx}, the singlet mass was around 50\%
lighter than the non-singlet scalar mass.

\begin{figure}[t]
%%%\begin{center}
\leavevmode
\includegraphics[scale=0.35,clip,angle=270]{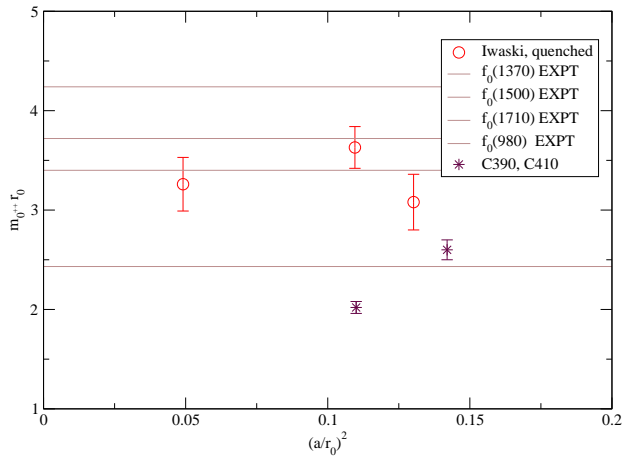}
%%%\end{center}
\caption[]{\label{fig:Iwasakimass} {
FS $0^{++}$ mass from this calculation compared
to the quenched $0^{++}$ glueball mass from 
Necco~\cite{Necco:2003vh}.
}}
\end{figure}

Necco~\cite{Necco:2003vh}  has recently compared the continuum limit 
for various quantities, including the mass of the  $0^{++}$ glueball
state, in quenched QCD using a number of improved gauge actions.
In figure~\ref{fig:Iwasakimass} we plot the masses obtained from
the C390 and C410 data sets with the quenched $0^{++}$ glueball masses
from Necco~\cite{Necco:2003vh}. This shows that a lower mass is obtained 
with dynamical quarks for this improved gauge action also.

\begin{figure}[t]
%%%\begin{center}
\leavevmode
\includegraphics[scale=0.35,clip,angle=270]{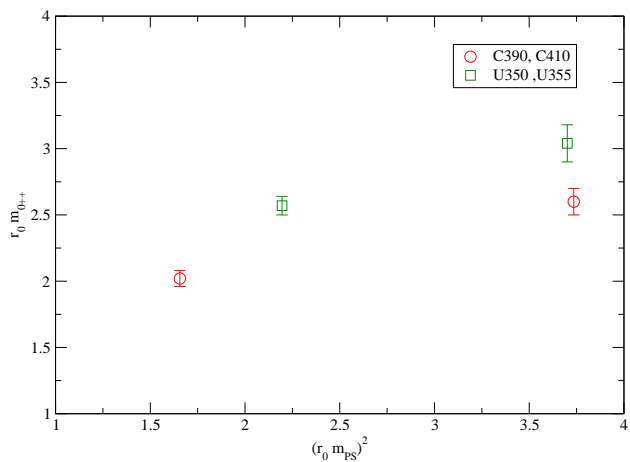}
%%%\end{center}
\caption[]{\label{fig:gmass} {
FS $0^{++}$ mass as a function of pion mass squared.
}}
\end{figure}

The plot of the $0^{++}$ mass as a function of the square of the
lattice spacing (figure~\ref{fig:contlimit}) 
is slightly misleading for unquenched QCD, because it
does not show the quark mass dependence. In figure~\ref{fig:gmass}
we plot the FS $0^{++}$ mass as a function of the pion mass squared.
We didn't attempt a chiral extrapolation.

We have not included in the summary figure~\ref{fig:contlimit},
the recent results from~\cite{Gregory:2005yr}. This calculation does
not see a deviation of the unquenched FS $0^{++}$ masses
(using glueball interpolating operators)
from the quenched $0^{++}$ glueball masses, but did not
include simultaneous fits to the glue and fermionic 
interpolating operators.

\section{Decay}

\begin{figure}[t]
%%%\begin{center}
\leavevmode
\includegraphics[scale=0.35,clip,angle=270]{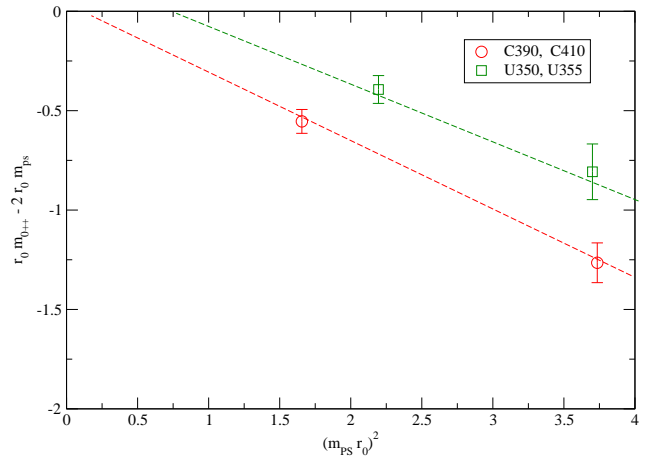}
%%%\end{center}
\caption[]{\label{fig:mgluePION} {
Two pion threshold for the FS scalar meson as a function of pion mass squared.
}}
\end{figure}

The dominant hadronic decay of the flavour-singlet $0^{++}$ meson is to
two pions. In figure~\ref{fig:mgluePION} we show the relevance of  the
decay threshold. The proximity of this threshold may have
significant impact on the spectrum, also it allows  an investigation of
the decay transition strength itself.

 The $f_0$ meson has an allowed S-wave decay to $\pi \pi$ and this has 
a low threshold so will have a significant impact on the analysis. In a 
finite spatial volume, as on a lattice, the two body states are discrete
 and one can concentrate on the lightest state, with the two pions at
rest.  A  study of the FS sector should then involve gluonic
operators (glueball-like);  fermionic ($\bar{q} q$) and two meson ($\pi
\pi$) to give the fullest coverage. We have studied the spectrum produced 
by the first two types of operator above. Here we discuss the prospects for 
including the two pion channel.
 
Historically the two pion channel was included in a quenched study of 
the scalar glueball by Weingarten and Sexton~\cite{Sexton:1995kd}
although their results  were very preliminary. More recently two body
channels have been  studied~\cite{McNeile:2002fh,McNeile:2006nv} in
connection with mesonic decays, such as $\rho \to \pi \pi$ and  $ a_0
\to K K$.  

In figure~\ref{fig:mgluePION} the threshold for scalar decay to two
lattice pions is plotted.
The U355 data has  lattice  $f_0$ states at 0.51(2) and
0.92(6) (in lattice units) compared to the $\pi \pi$ threshold on that 
lattice at 0.59. Thus the heavier $f_0$ is unstable. This implies that
the  impact of the two pion channel could potentially be strong.

The fact that the $f_0$  and two pion states are close means that we may
be able to compute decay widths, as was attempted by the GF11
group~\cite{Sexton:1995kd}, using the formulation  developed  for the
decay of the heavy $1^{-+}$ hybrid~\cite{McNeile:2002az}
 and $\rho$ meson~\cite{McNeile:2002fh,Michael:2005kw}.

Of course the theoretical problem of dealing with unstable
particles has been solved in principle by 
L\"{u}scher over ten years ago~\cite{Luscher:1991cf}. In his formalism, 
the scattering phase shift can be extracted from the volume
dependence of two-body energy levels. The mass and width of the resonance
can then be extracted from the phase shift. Sufficient precision has not been 
available to pursue this, so far. Thus the more qualitative treatment of 
hadronic transitions,  described above, has been used, instead.

\begin{figure}
  \includegraphics[width=.8\linewidth]      {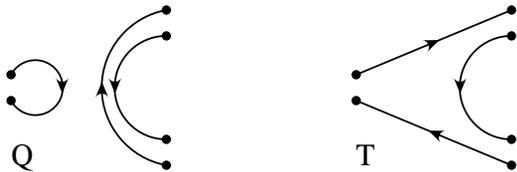}
 %\vbox{\epsfxsize=4in \epsfbox{fd_qt.ps}}
 \caption{The quark structure of the disconnected contribution ($Q$) 
and the triangle contribution ($T$) to the transition $f_0 \to \pi \pi$.
}
\label{fig:fd_qt}

\end{figure}

 The lack of progress in the study of $f_0$ meson decay to $\pi \pi$  is
due to the presence of  disconnected diagrams in the decay- see 
fig.~\ref{fig:fd_qt}.
 Following the methods used for $\rho$ decay, one can estimate the
relative importance of  this disconnected contribution $Q$ compared to
the connected triangle diagram $T$. The triangle contribution is
relatively easy  to evaluate unlike the disconnected contribution. In 
fact results for the transition $T$ of a scalar to two pseudoscalar mesons
have been presented~\cite{McNeile:2006nv} elsewhere.

\begin{figure}
\vbox{\epsfxsize=4in \epsfbox{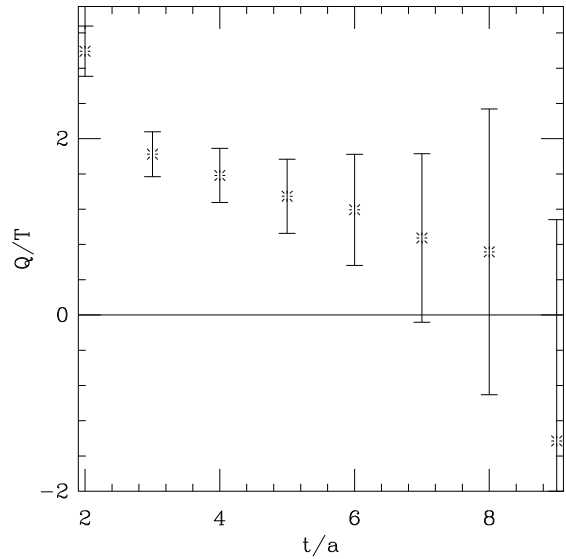}}
 \caption{The relative contribution of the disconnected contribution
($Q$) (vacuum subtracted)  and the triangle contribution ($T$) to the
transition $f_0 \to \pi \pi$ with  all states at zero momentum for
lattice U355. } \label{fig:qt}

\end{figure}

 Using time-slice stochastic sources (as used by
ref.~\cite{McNeile:2002fh}) from 100 gauge configurations of U355,  and
combining these with the  volume stochastic source results used above,
we can evaluate these contributions: see fig.~\ref{fig:qt}. This 
illustrates that even though we have measured the disconnected
contribution $Q$ from every  time-value and space-value on each of 100
gauge configurations, the resulting  error is very large. 

 Thus a thorough study of decays of FS scalar mesons must await a  study
with much larger  statistics than that we have available.

\section{Discussion and conclusions} \label{se:conclusions}

The inclusion of the $\overline{\psi} \psi$ $0^{++}$ 
operators with the 
pure glue $0^{++}$ operators in the 
variational analysis has produced a suppression of the masses 
relative to the quenched results at an equivalent lattice 
spacing (see figure~\ref{fig:contlimit}). There are two main 
possible interpretations
of these results. It could be possible that the lattice errors ($O(a^2)$)
in the FS $0^{++}$ masses are larger for unquenched QCD than for 
quenched QCD, because of the phase structure of the theory.
Another possibility is that in the unquenched calculations the FS $0^{++}$
mesons couple to the $f_0(980)$ or even to the $f_0$(400-1200), thus driving 
their mass way below the quenched $0^{++}$ glueball mass at 1600 MeV. We
discuss these two possibilities in turn.

In quenched QCD the $O(a^2)$ lattice spacing errors to the 
mass of the $0^{++}$ glueball are large (see the quenched data 
in figure~\ref{fig:contlimit}). There are lattice formalisms 
that add irrelevant operators to the lattice actions that 
in principle can remove the $O(a^2)$ errors (this is 
known as improvement). The first systematic use of the improvement
formalism for the $0^{++}$ glueballs was by 
Morningstar and Peardon~\cite{Morningstar:1997ff,Morningstar:1999rf}.
They also found large lattice spacing dependence in the 
mass of the $0^{++}$ glueball even with their improved gauge 
actions and this phenomena is colloquially known in lattice QCD
circles as
"the scalar dip". Further studies of the $0^{++}$ 
masses on the lattice with a variety of improved gauge actions
by Necco~\cite{Necco:2003vh} and 
Niedermayer et al.~\cite{Niedermayer:2000yx,Rufenacht:2001pi}
still found a strong lattice dependence of the results.
The lattice spacing dependence of the $0^{++}$ mass is 
probably the best known example of the break down of 
the simplest application of the improvement and perfect action
programs to reduce lattice spacing dependence.

In quenched  QCD the explanation of the strong lattice spacing
dependence of the mass of the $0^{++}$ state is the influence
of the bulk phase transition in the adjoint 
plane~\cite{Heller:1995bz,Blum:1994xb} close to the region where
calculations are done. The way to reduce the lattice spacing dependence 
is to add an adjoint term to the gauge action with a negative
coefficient to work in a region of the parameter space 
away from the phase transition~\cite{Morningstar:1998du,Hasenbusch:2004yq}.
This matters for unquenched calculations because it has been conjectured
that the clover fermion action induces an adjoint gauge term
that will increase the lattice spacing dependence of the 
FS $0^{++}$ mesons~\cite{Hart:2001tv,Sommer:2003ne,Aoki:2004iq}. If this 
interpretation
is correct, then the suppression of the FS $0^{++}$ masses in
figure~\ref{fig:contlimit} is purely a lattice artifact.

There is additional evidence that suggests that the 
suppression of the masses of FS $0^{++}$ mesons 
seen in this calculation is just not a lattice artifact.
The adjoint term is also thought to be the cause of the 
phase transition seen by JLQCD~\cite{Aoki:2004iq} and 
Farchioni et al.~\cite{Farchioni:2004fs}. The phase structure
of the unquenched lattice calculations that use Wilson
like fermion actions have been 
reviewed by Shindler~\cite{Shindler:2005vj}. 
Both JLQCD~\cite{Aoki:2004iq} and Farchioni et al.~\cite{Farchioni:2004fs}
found that the phase transition was weakened or removed by the 
use of an improved gauge action, such as the Iwasaki action. 
The consistency of the suppression of the FS $0^{++}$ masses
between calculations that use the Wilson and Iwasaki gauge actions in 
figure~\ref{fig:contlimit} seems to us good evidence
that the suppression of the masses of the FS $0^{++}$ mesons
is not just a lattice artifact.

There are other physical reasons why intuition based 
on the phase structure of quenched QCD may be not be a good 
guide for unquenched lattice calculations of the FS $0^{++}$
mesons. If the $f_0$(980) or $f_0$(400-1200) have any 
quark - anti-quark components then they will couple to our 
interpolating operators and drive the ground state 
of the $0^{++}$ channel to the 1 GeV level. Although there 
have been recent claims by Mathur et al.\cite{Mathur:2006bs}
that the $a_0$(980) is molecular, an unquenched calculation
by UKQCD provided evidence for the $a_0$(980) state to be a
quark - anti-quark  state~\cite{McNeile:2006nv}.

One of the advantages of lattice QCD calculations is that they can
be used to understand the physical mechanisms behind the numbers.
In the quenched theory the $\overline{q}q$ and glueball states couple
to distinct $0^{++}$ states. Calculations that include very heavy
dynamical quarks should be similar to the distinct quenched states.
As the mass of the sea quarks is reduced the states will start to mix.

For example: taking $a_0$ masses from ref.~\cite{McNeile:2006nv}, we get
 at U355  $r_0 m(a_0)=3.23(20)$. Then at this lattice spacing,  the
quenched glueball mass is around $r_0 m(GB)=3.6$. A mixing shift of $r_0
\Delta E= 0.63$ would move these levels to 2.6 and 4.2 respectively,  in
excellent agreement with our two observed levels at 2.57(10) and
4.63(30)   for U355. Furthermore, since the mixing shift  is large
compared to the initial splitting (0.37 in the above example),  we would
expect approximately maximal mixing in the observed spectrum -  which is
indeed what we find from the fit coefficients. Defining a mixing from
 $$ \tan^2 \alpha = 
      -{\langle 0 | G | 1 \rangle  \langle 0 | \bar{q}q | 2 \rangle 
  \over \langle 0 | \bar{q}q | 1 \rangle \langle 0 | G | 2 \rangle} 
 $$
 where $G$ and $\bar{q}q$ refer to the operators used to create the 
states and 1 and 2 refer to the two lightest states observed on the
lattice. The $\langle 0 | G | 1 \rangle$ are one of the 
$c_{j}^{n}$ coefficients in equation~\ref{eq:matrixFIT}.
For U355 with momentum $n.n=1$, we obtain a mixing angle
$\alpha=57(10)^0$ where $45^0$ would correspond to  maximal mixing.

 % and this value corresponds to the lattice ground state being slightly 
 % more gluonic than fermionic.

 In terms of a mass mixing matrix, this example would correspond to
 \begin{equation}
 \left( \begin{array}{ll}
 r_0 m(\bar{q}q) &  r_0 X \\
 r_0 X  & r_0 m(GB)
 \end{array}  \right) =
 \left( \begin{array}{ll}
 3.2 &  0.8 \\
 0.8  & 3.6
 \end{array}  \right) 
 \end{equation} % fix trace and det to be same -> 2.6 4.2
 so that the mixing matrix element is $r_0 X=0.8$ and the resulting 
eigenvalues are those we observe (2.6 and 4.2) on the lattice. The 
mixing would be nearly maximal (actually $38^0$) in this example.
 This same mixing matrix element ($X$) and glueball mass with $r_0
m(\bar{q}q)=3.56(14)$~\cite{McNeile:2006nv}  for U350 would yield
eigenstates at 2.8 and 4.4 where the former is in  good agreement with
xour lattice result, 2.85(19), for U350.
 For the C410 data, at a coarser lattice spacing, a larger value of $r_0
X \approx 1.1$ would give  a splitting of the $r_0 m(\bar{q}q) = 3.1(1)$
and $r_0 m(GB) \approx 3.1$  values to agree with our lightest $f_0$
state at $r_0 m_1=2.0(1)$. 
 In an earlier study with $N_f=2$ at a coarser lattice spacing,  an
estimate of $r_0 X \approx $ 1.5 
($3.65 \sqrt{2}\ 0.3 $)  
was quoted~\cite{McNeile:2000xx}.

Thus we find that in our lattice spacing and quark mass range, the 
fermionic (with $N_f=2$ degenerate sea-quarks) and gluonic operators
create states which  are maximally mixed, with a mixing matrix element
of $r_0 X \approx 0.8$ which corresponds to an energy of approximately 
320~MeV. This value is smaller than that estimated previously using
$N_f=2$  sea-quarks with a coarser lattice spacing and  than that  we
estimate from C410 with a similar coarse lattice spacing.

If the mixing matrix element $X$ were to remain constant as the quark
mass  is reduced, then the estimate~\cite{McNeile:2006nv} of the $a_0$
mass (for $N_f=2$) is 1 GeV  while the glueball is around 1.6 GeV. These
input masses would then be mixed to 0.86 and 1.74 GeV. This gives some
indication of the  magnitude of mixing effects which could be present in
the experimental spectrum and require calculation with similar
small lattice spacings.

In the experimental spectrum, however, there are expected to 
be significant effects arising from the $\bar{s}s$ scalar meson, which 
we have neglected here. This $\bar{s}s$ state should mix with gluonic 
operators and one will need a $3 \times 3$ model to incorporate this 
adequately. Lattice QCD is able to include 2+1 flavours of sea-quark and 
this will be appropriate for a fuller confrontation with experiment.

%%
%%  the future
%%  Continuum limit, hard O(a^2) improved action
%%  lighter quark masses
%%

In quenched QCD the  first calculation~\cite{Bali:1993fb} that  computed
the mass of the $0^{++}$ glueball  in the continuum limit used a lattice
spacings down to 0.05~fm to have a reliable continuum extrapolation.
Unquenched calculations that use the Wilson gauge action may also have
similar scaling violations. 

As we have noted, there appears to be a significant lattice spacing
dependence in the mixing.  To quantify the amount of glueball to
$\overline{q}q$ mixing in $f_0$ mesons will   require a dynamical
fermion calculation at even finer lattice spacings such as 0.05 fm.  For
Wilson type fermions, recent algorithmic 
advances~\cite{Hasenbusch:2001ne,Urbach:2005ji,Luscher:2005mv}
means that this may
now be just attainable on
the current generation of machines.

 The eventual definitive study of flavour singlet scalar mesons will
need  fine lattice spacing, dynamical simulations with light quarks
(plus a strange quark in the sea) and  large statistics to enable the 
disconnected diagrams to be evaluated accurately. The methods we have 
used with $N_f=2$ degenerate sea-quarks of mass down to about 50\% of
the  strange quark mass show the way forward. We find substantial mixing
of  glueballs with $\bar{q}q$ states in the $f_0$ spectrum and we expect
that feature  to survive in a future study.

Although, given the qualifications we mention above about this 
calculation, the message we get from the summary of glueball 
masses in figure~\ref{fig:contlimit}
is that the effect of unquenched fermions is to drive the 
mass of the FS state from the 1600 MeV of the quenched glueball
towards 1 GeV. This calculation does not favour the weak
mixing of glue and $\overline{\psi}\psi$ operators
proposed by Weingarten and Lee and 
other groups~\cite{Lee:1999kv,Cheng:2006hu}. Looking at our data
it is not clear that a mixing scheme based on only the 
states $f_0(1370)$, $f_0(1500)$, and $f_0(1710)$ 
is complete enough to determine the fate of the 
quenched glueball.

%%%%%%%%%%%%%%%%%%%%%%%%%%%%%%%%%%%%%%%%%%%%%%%%%%%%
%%           References                          %%%
%%%%%%%%%%%%%%%%%%%%%%%%%%%%%%%%%%%%%%%%%%%%%%%%%%%%

\section{Acknowledgements}

The authors acknowledge support from  PPARC grant PPA/Y/S/2003/00176.
A.H. thanks the U.K. Royal Society for financial
support.
 This work has been supported in part by the EU Integrated
 Infrastructure Initiative Hadron Physics (I3HP) under contract
  RII3-CT-2004-506078.
We are grateful to the ULgrid project of the
University of Liverpool for computer time.
 We acknowledge the CP-PACS collaboration~\cite{AliKhan:2001tx} for
making available  their gauge  configurations.

%%\bibliographystyle{h-physrev2}
%%\bibliography{conferences,fermi_glue,fa0,lhyb} 

\pagebreak
%%%
%%%
%%%

\end{document}